\documentclass[aps,showpacs,twocolumn,superscriptaddress]{revtex4-2}
\usepackage[utf8]{inputenc}
\usepackage{amsfonts}
\usepackage{amssymb}
\usepackage{amsmath}
\usepackage{graphicx}
\usepackage{epsfig}
\usepackage{subfigure}
\usepackage{appendix}
\usepackage{hyperref}
\usepackage{color}

\hypersetup{hypertex=true, colorlinks=true, linkcolor=blue, urlcolor=blue, citecolor=blue}

\begin{document}

\title{Dynamical spin-nematic correlation in a transverse field Ising chain with non-Hermitian Gamma interaction}
\author{Yu-Hong Yan}
\affiliation{School of Physics and Optoelectronic Engineering, Foshan
University, Foshan, 528225, China}
\author{Ran Wang}
\affiliation{College of Physics and Materials Science, Tianjin Normal University,Tianjin 300387, China}
\author{Kun-Liang Zhang}
\email[Contact author: ]{zhangkl@fosu.edu.cn}
\affiliation{School of Physics and Optoelectronic Engineering, Foshan
University, Foshan, 528225, China} 
\affiliation{Guangdong-HongKong-Macao Joint Laboratory for Intelligent Micro-Nano Optoelectronic Technology, Foshan 528225, China}
\date{\today}

\begin{abstract}
We investigate the effect of non-Hermitian Gamma interaction on the phase transitions and magnetic correlations for the transverse field Ising chain. We demonstrate that apart from the gapped antiferromagnetic and paramagnetic phases, there is a gapless phase induced by parity-time symmetry breaking, where the system exhibits long-range and short-range spin-nematic correlations in different regions divided by the quantum critical line determined from the correlation function and the subsystem entanglement entropy. Furthermore, we reveal that the parity-time symmetry breaking leads to the emergence of dynamical spin-nematic correlation, which also suggests a way of characterizing the spin-nematic map through non-equilibrium dynamics.  Our findings show rich quantum phases stem from the competition among the Ising interaction, transverse field and non-Hermitian Gamma interaction, as well as providing a scheme for generating spin-nematic correlation in the spin chain.

\end{abstract}

\maketitle

\section{Introduction}
\label{introduction}

Quantum spin chains provide an ideal platform for investigating quantum phase transitions (QPTs) \cite{Sachdev1999, Pfeuty1970, coldea2010quantum}, symmetry-protected topological order \cite{haldane1983nonlinear, affleck1987rigorous, pollmann2012symmetry, chen2011classification}, and fractionalized excitations \cite{tennant1993measurement, giamarchi2004quantum, mourigal2013fractional} due to their strong quantum fluctuations and exact solvability. Traditional spin-chain models, such as the Heisenberg chain, mainly focus on isotropic interactions. Nevertheless, in magnetic materials containing heavy transition-metal elements, strong spin-orbit coupling breaks the spin rotational symmetry and gives rise to highly bond-dependent anisotropic interactions \cite{jackeli2009mott}. In this regard, beyond the well-known Kitaev interaction \cite{rau2014, winter2017models, Yang2021, takahashi2021topological, luo2023spontaneous}, the symmetric off-diagonal exchange, namely the Gamma interaction with the form $\mathit{\Gamma}\sum_{j}(\hat{\sigma}_j^x \hat{\sigma}_{j + 1}^y + \hat{\sigma}_j^y \hat{\sigma}_{j + 1}^x )$, plays a particularly important role \cite{Andreev1984, Penc2010, lu2014nematic, gohlke2018quantum, Soerensen2021, Zvyagin2022, Zvyagin2023, Luo2024, zvyagin2024intersite, Abbasi2025, zvyagin2025intersite, dai2026phase}. Introducing the Gamma interaction into quantum spin chains leads to strong energetic competition and pronounced quantum frustration, which may not only suppress conventional long-range magnetic order, but also induce exotic quantum many-body states, including quantum spin liquids \cite{gohlke2018quantum}, spin-nematic orders \cite{Soerensen2021, Luo2024, Abbasi2025}, and topological excitations \cite{yang2020phase}. Therefore, the extension from conventional spin chains to more complex spin-chain models incorporating the Gamma interaction marks a deeper understanding of quantum magnetic materials.

Recent studies reveal that dissipation can be utilized to modulate chiral spiral phases in the non-Hermitian Ising-Gamma model \cite{Huang2026}, while also driving exotic correlation spreading and entanglement generation in non-Hermitian spin chains \cite{Turkeshi2023}. Traditionally, dissipation has long been regarded as a negative factor to be minimized or avoided. Non-Hermitian physics \cite{ashida2020non}, however, turns this paradigm on its head by transforming dissipation into a fundamentally new degree of freedom for system control. By precisely tailoring gain and loss, non-Hermiticity is demonstrating irreplaceable value in directional wave control \cite{lin2011unidirectional, jin2018incident, weidemann2020topological, jin2021symmetry}, the engineering of novel topological states \cite{kawabata2019symmetry, wu2020floquet, wu2025topological}, the development of novel light sources utilizing topological lasers \cite{feng2014single,hodaei2014parity,bandres2018topological}, and the design of quantum devices through dissipation engineering \cite{verstraete2009quantum,naghiloo2019quantum}. In recent years, the research interest of non-Hermitian physics has expanded into the field of quantum spin systems, both theoretically \cite{Lee2014, li2016chern, zhang2020dynamic, Zhang2020, Liu2021, Zhang2021a, Lenke2021, Matsumoto2022, Guo2022, Sun2022a, Lenke2023, zhang2024magnetic, zhang2024magnetization} and experimentally \cite{naghiloo2019quantum, Partanen2019, Wu2019a, Li2019a, Zhang2021, Ren2022, gao2024experimental}, demonstrating that non-Hermitian interactions can drive QPTs, give rise to exotic dynamical behaviors and significantly alter the phase diagram of the system. Consequently, the role of non-Hermitian interactions in unconventional magnetic orders warrants further research, particularly in terms of their non-equilibrium dynamics.

The present work is devoted to investigating the effect of non-Hermitian Gamma interaction on the QPT and magnetic correlations for the transverse field Ising chain. Using the free fermion technique, we are able to obtain magnetic correlations, dynamical characteristics and phase diagram of the chain exactly. We show that apart from the gapped antiferromagnetic and paramagnetic phases, there is a non-Hermitian gapless phase induced by parity-time (PT) symmetry breaking when increasing the strength of non-Hermitian Gamma interaction, where the system exhibits long-range and short-range spin-nematic correlations in different regions divided by the quantum critical line determined from the correlation function and the subsystem entanglement entropy. This stands in stark contrast to the absence of true long-range order in the gapless phases of Hermitian spin chains. Furthermore, we reveal that the PT symmetry breaking leads to the emergence of dynamical spin-nematic correlation, which also suggests a way of characterizing the spin-nematic map through non-equilibrium dynamics.

The rest of this paper is organized as follows: In Sec. \ref{model}, we introduce the Hamiltonian of the transverse field Ising chain with non-Hermitian Gamma interaction,  and diagonalize the Hamiltonian and obtain the phase diagram defined by the energy gap. In Sec. \ref{order_parameters},  we discuss the behaviors of the spin-$xx$, the spin-nematic correlations and the subsystem entanglement entropy. In Sec. \ref{dynamical_nematics_order}, we investigate the dynamics of spin-nematic correlation, and reveal the spin-nematic map through non-equilibrium dynamics. Finally, we discuss and conclude our findings in Sec. \ref{conclusion}.

\section{Model and diagonalization}
\label{model}

We initiate our study by introducing the model Hamiltonian of the transverse field Ising chain with non-Hermitian Gamma interaction, which has the form 
\begin{equation}
\begin{aligned}
  \hat{\mathcal{H}} =& J \sum^N_{j = 1} \hat{\sigma}_j^x \hat{\sigma}_{j + 1}^x + h \sum^N_{j =
  1} \hat{\sigma}_j^z\\
   &- \mathrm{i} \mathit{\Gamma} \sum^N_{j = 1} (\hat{\sigma}_j^x \hat{\sigma}_{j + 1}^y + \hat{\sigma}_j^y \hat{\sigma}_{j + 1}^x ),
   \end{aligned}
   \label{H_NH}
\end{equation}
where $\hat{\sigma}_{j}^{x,y,z}$ are Pauli operators on site $j$. In the above Hamiltonian, the first term describes the antiferromagnetic Ising interaction with $J>0$, the second term corresponds to the transverse field with strength $h$, and the last term represents imaginary symmetric off-diagonal Gamma interaction with strength $\mathit{\Gamma}$. The length of the chain is $N$ and we assume periodic boundary condition, that is $\hat{\sigma}_{N+1}^{x,y}=\hat{\sigma}_{1}^{x,y}$. 

In general, dissipation and decoherence are unavoidable in practical quantum systems. While these effects are traditionally described using the quantum master equation, non-Hermitian Hamiltonians are primarily understood as an effective framework for describing such open quantum systems \cite{ashida2020non}. The imaginary symmetric off-diagonal Gamma interaction occurs in the no-click limit of the stochastic quantum jump trajectories when correlated jump operators $ \sqrt{\mathit{\Gamma}} (\hat{\sigma}_j^x + \hat{\sigma}_{j + 1}^y)$ and $ \sqrt{\mathit{\Gamma}} (\hat{\sigma}_j^y + \hat{\sigma}_{j + 1}^x)$ are measured, in which $\mathit{\Gamma}\geqslant 0$ is the dissipation rate \cite{Daley2014, Turkeshi2023} (see \ref{appendix_a}). Without loss of generality, we also assume $h\geqslant 0$ in the subsequent discussions.

When $\mathit{\Gamma}= 0$, the model in Eq. (\ref{H_NH}) is the celebrated transverse field Ising chain \cite{Pfeuty1970}, which respects parity symmetry with the parity operator being $\hat{\mathcal{P}}=\prod_{j=1}^{N}\hat{\sigma}_{j}^{z}$. This is a $\mathbb{Z}_2$ symmetry intimately connected with the QPT from a parity-symmetric paramagnetic phase to an antiferromagnetic phase with spontaneously broken parity symmetry. We note that the non-Hermitian Gamma interaction does not break the parity symmetry, we have $\hat{\mathcal{P}}\hat{\mathcal{H}}\hat{\mathcal{P}}^{-1}=\hat{\mathcal{H}}$ when $\mathit{\Gamma}\neq 0$, and $\hat{\mathcal{H}}$ also preserves time-reversal symmetry $\hat{\mathcal{T}}\hat{\mathcal{H}}\hat{\mathcal{T}}^{-1}=\hat{\mathcal{H}}$ with $\hat{\mathcal{T}}$ being the complex conjugation operator: $\hat{\mathcal{T}}\mathrm{i}\hat{\mathcal{T}}^{-1}=-\mathrm{i}$, $\hat{\mathcal{T}}(\hat{\sigma}_{j}^{x},\hat{\sigma}_{j}^{y},\hat{\sigma}_{j}^{z})\hat{\mathcal{T}}^{-1}=(\hat{\sigma}_{j}^{x},-\hat{\sigma}_{j}^{y},\hat{\sigma}_{j}^{z})$ . As a result, the non-Hermitian Hamiltonian $\hat{\mathcal{H}}$ respects PT symmetry, that is $\hat{\mathcal{P}}\hat{\mathcal{T}}\hat{\mathcal{H}}(\hat{\mathcal{P}}\hat{\mathcal{T}})^{-1}=\hat{\mathcal{H}}$. Note that in spin lattices, the definition of the $\hat{\mathcal{P}}\hat{\mathcal{T}}$ operator is not unique \cite{CastroAlvaredo2009}. In non-Hermitian systems, PT symmetry does not require the system to be individually invariant under parity and time-reversal operations. Indeed, in most situations, PT-symmetric non-Hermitian systems are invariant under neither parity nor time reversal alone. Interestingly, in our case, the parity symmetry of $\hat{\mathcal{H}}$ implies the preservation of Ising-type phase transition, and the PT symmetry may lead to a full real spectrum and non-Hermitian phase transition \cite{Bender1998, Mostafazadeh2002}. Therefore, the phase transition and magnetic correlations are governed by the competition between PT symmetry breaking and spontaneous parity symmetry breaking.

 \begin{figure*}[tbh]
\centering
\includegraphics[width=1\textwidth]{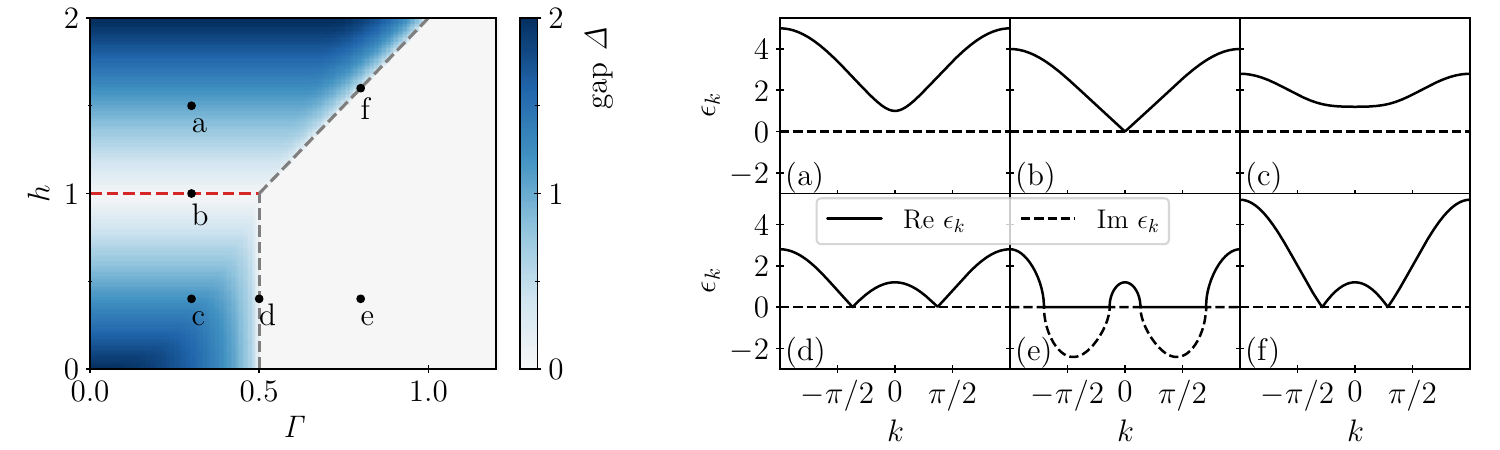}
\caption{Left panel is the phase diagram characterized by the energy gap $\mathit{\Delta}$, as a function of $\mathit{\Gamma}$ and $h$. The dashed lines represent the phase transition lines, where the red dashed line denotes that of Ising-type phase transition, and the gray dashed lines denote that of non-Hermitian phase transition separating the PT-symmetric and PT-broken phases. Right panels show the dispersion $\epsilon_{k}$ for several representative points in parameter space of $(\mathit{\Gamma},\ h)$, which are marked in phase diagram of left panel and taken as (a) $(0.3,\ 1.5)$ in gapped phase; (b) $(0.3,\ 1.0)$ in phase boundary with a critical mode $k_{\mathrm{c}}=0$; (c) $(0.3,\ 0.4)$ in gapped phase; (d) $(0.5,\ 0.4)$ in phase boundary with two critical modes; (e) $(0.8,\ 0.4)$ in gapless phase; (f)  $(0.8,\ 1.6)$ in phase boundary with two critical modes. We note that in the gapless phase (white region), the dispersion $\epsilon_{k}$ become complex and appear in complex-conjugate pairs. In all situations, we take $N=2000$ and $J=1$.}
\label{fig_gap_phase_diagram}
\end{figure*}

The Hamiltonian in Eq. (\ref{H_NH}) can be solved by Jordan-Wigner fermionization, followed by free-fermion diagonalization (see \ref{appendix_b}). Explicitly, through the Jordan-Wigner transformation \cite{jordan1928paulische}
\begin{equation}
\begin{aligned}
	\hat{\sigma}_j^x &= \prod_{l < j} (1 - 2 \hat{c}_l^{\dagger} \hat{c}_l) (\hat{c}_j^{\dagger} +
  \hat{c}_j),\\
  \hat{\sigma}_j^y &= \mathrm{i} \prod_{l < j} (1 - 2 \hat{c}_l^{\dagger} \hat{c}_l) (\hat{c}_j^{\dagger} -
  \hat{c}_j),\\
  \hat{\sigma}_j^z &= 1 - 2 \hat{c}_j^{\dagger} \hat{c}_j,
  \end{aligned}
  \label{J-W_transformation}
\end{equation}
and the Fourier transformation
\begin{equation}
  \hat{c}_j =  \frac{e^{- \mathrm{i} \pi / 4}}{\sqrt{N}} \sum_k e^{\mathrm{i} k j} \hat{c}_k,
\end{equation}
the Hamiltonian can be written as the Bogoliubov-de Gennes (BdG) form
\begin{equation}
	 \hat{\mathcal{H}} =  \sum_k \left(\begin{array}{cc}
    \hat{c}_k^{\dagger} & \hat{c}_{- k}
  \end{array}\right) \mathcal{H}_k \left(\begin{array}{c}
    \hat{c}_k\\
    \hat{c}_{- k}^{\dagger}
  \end{array}\right),
  \label{H_BdG}
\end{equation}
where
\begin{equation}
	  \mathcal{H}_k =  \left(\begin{array}{cc}
    J \cos k - h & - J  \sin k - 2\mathit{\Gamma} \sin k\\
    - J \sin k + 2 \mathit{\Gamma} \sin k & - J \cos k + h
  \end{array}\right).
\end{equation}

Then the Hamiltonian can be diagonalized as 
\begin{equation}
	\begin{aligned}
		\hat{\mathcal{H}} = & \sum_k \epsilon_k \left( \overline{\hat{\eta}}_k
  \hat{\eta}_k - \frac{1}{2} \right),\\
  \epsilon_k = & \pm 2 \sqrt{(J \cos k - h)^2 + (J^2 - 4 \mathit{\Gamma}^2) \sin^2 k},
	\end{aligned}
	\label{H_diagonalization}
\end{equation}
through the complex Bogoliubov transformation
\begin{equation}
	\begin{aligned}
		\hat{\eta}_k = & \mu_k \hat{c}_k + \nu_k e^{\lambda} \hat{c}_{- k}^{\dagger},\\
  \overline{\hat{\eta}}_k = & \mu_k \hat{c}_k^{\dagger} + \nu_k e^{- \lambda} \hat{c}_{- k},
	\end{aligned}
	\label{Bogoliubov_transformation}
\end{equation}
where
\begin{equation}
	\begin{aligned}
		\mu_k = &  \sqrt{\frac{\epsilon_k / 2 + J \cos k - h}{\epsilon_k}},\\
  \nu_k = & -\sqrt{\frac{(J^2 - 4 \mathit{\Gamma}^2) \sin^2 k}{\epsilon_k (\epsilon_k / 2
  + J \cos k - h)}},\\
  e^{\lambda} = & \sqrt{\frac{J + 2 \mathit{\Gamma}}{J - 2 \mathit{\Gamma}}},
	\end{aligned}
\end{equation}
and $\hat{\eta}_k$, $\overline{\hat{\eta}}_k$ obey the following canonical anticommutation relation
\begin{equation}
	\begin{aligned}
		\left\{ \hat{\eta}_k,  \overline{\hat{\eta}}_{k^{\prime}}\right\}=&\delta_{k,k^{\prime}},\\
		\left\{ \hat{\eta}_k,  \hat{\eta}_{k^{\prime}}\right\}=&\left\{  \overline{\hat{\eta}}_k,  \overline{\hat{\eta}}_{k^{\prime}}\right\}=0.
	\end{aligned}
\end{equation}
Note that $\overline{\hat{\eta}}_k \neq \hat{\eta}_{k}^{\dagger}$ for the non-Hermitian case $\mathit{\Gamma}\neq 0$. 

Here and in what follows, we focus on the properties of the system's ground state $\left|\mathcal{G} \right\rangle$, which fulfills $\hat{\eta}_{k}\left|\mathcal{G} \right\rangle=0$ and is also called Bogoliubov vacuum. Then up to a normalization factor, the ground state has the form $\left|\mathcal{G} \right\rangle=\prod_{k}\hat{\eta}_{k}\left|0 \right\rangle$, where $\left|0 \right\rangle$ is the vacuum for the fermions, that is $\hat{c}_{m}\left| 0 \right\rangle=0$. According to the diagonalization form in Eq. (\ref{H_diagonalization}), it can be checked that the ground-state energy is
\begin{equation}
	\mathcal{E}=-\frac{1}{2}\sum_{k}\epsilon_{k}.
\end{equation}
For the dispersion $\epsilon_{k}$ in Eq. (\ref{H_diagonalization}), one is free to choose either the “$+$” or the “$-$” branch \cite{Lee2014}. In the PT-symmetric region, we choose the  “$+$”  branch so that the ground-state energy $\mathcal{E}$ is minimal. While in the PT-broken region, $\epsilon_{k}$ is not always real, then for each $k$ with complex $\epsilon_{k}$, we fix the sign convention so that $\operatorname{Im}\epsilon_{k} <0 $. 
We emphasize that while in the PT-symmetric region the state $ |\mathcal{G}\rangle$ can be understood as ground state, it should be understood as a reference state selected by the non-Hermitian spectral structure and the associated dynamics in the PT-broken region. In our treatment, the branch with $\operatorname{Im}\epsilon_{k} <0 $ is chosen because, under the non-Hermitian time evolution generated by $\exp(-\mathrm{i}\hat{\mathcal{H}}t)$, this branch determines the dynamically stable sector that remains physically relevant at long times. Thus, it provides a natural reference state for characterizing static correlation properties within the dynamical framework in Sec. \ref{dynamical_nematics_order}.

In general, the ground-state energy characterizes the thermodynamic non-analyticity of a QPT, while the excitation gap reveals critical low-energy excitations and typically closes at phase transition point or in the gapless phase. For a Hermitian or non-Hermitian system with full real $\epsilon_{k}$, the energy gap $\mathit{\Delta}$ can be denoted as the minimal values of $\epsilon_{k}$. However, for the PT-broken phase of the system, $\epsilon_{k}$ is not full real for all $k$. In non-Hermitian systems, energy gaps are categorized into two types, i.e., line gaps and point gaps, depending on whether the energy spectrum excludes a specific line or a single point in the complex energy plane \cite{kawabata2019symmetry}.  Here we defined the energy gap $\mathit{\Delta}$ as the minimal values of $\operatorname{Re}\epsilon_{k}$. Figure \ref{fig_gap_phase_diagram} shows the phase diagram characterized by the energy gap $\mathit{\Delta}$ in the $\mathit{\Gamma}$-$h$ plane (left panel) and the dispersion $\epsilon_{k}$ for several representative parameters (right panel). The phase diagram consists of two gapped phases and one gapless phase separated by three critical lines, along which the gap vanishes, i.e., $\epsilon_{k}=0$. The critical lines are given as follows: (i) $h=J \ (\mathit{\Gamma}<J/2)$ with critical mode $k_{\mathrm{c}}=0$; (ii) $\mathit{\Gamma}=J/2 \ (h<J)$ with critical modes $k_{\mathrm{c}}=\pm\arccos (h/J)$; and (iii) $h=2\mathit{\Gamma}\ (\mathit{\Gamma}>J/2)$ with critical modes $k_{\mathrm{c}}=\pm\arccos (J/h)$. For the typical  transverse field Ising chain  with $\mathit{\Gamma}=0$, it is clear that the paramagnetic phase and antiferromagnetic phase are separated by a critical point $h=J$, which extends to a critical line shown as a red dashed line of the phase diagram in Fig. \ref{fig_gap_phase_diagram} for the non-Hermitian case. To further clarify the nature of the reference states in these three phases, we will analyze the spin-correlation functions and spin-nematic indicator in the next section.

\section{Correlation functions and quantum criticality}
\label{order_parameters}

\subsection{Spin-correlation functions}
In this section, we evaluate the spin-correlation functions and spin-nematic indicator of the reference state $\left|\mathcal{G} \right\rangle=\prod_{k}\hat{\eta}_{k}\left|0 \right\rangle$. Before proceeding, it is useful to rewrite $\hat{\eta}_{k}$ in a more convenient form.  We point out that the complex Bogoliubov transformation in Eq. (\ref{Bogoliubov_transformation}) essentially serves as a similarity transformation for the biorthogonal diagonalization, in which the biorthogonal normalization breaks down at the exceptional points. Nevertheless, the solution of $\hat{\eta}_k$ derived without enforcing biorthonormal normalization remains valid over the entire parameter space. To facilitate the discussion that follows, we replace $\hat{\eta}_k$ with a solution without biorthogonal normalization, that is 
\begin{equation}
	\hat{\eta}_k =  u_k \hat{c}_k + v_k  \hat{c}_{- k}^{\dagger},
	\label{eta_uv}
\end{equation}
where
\begin{equation}
	\begin{aligned}
u_k=& J \cos(k)-h+\epsilon_k/2, \\
v_k=& -(J+ 2 \mathit{\Gamma})\sin(k).
	\end{aligned}
	\label{ukvk}
\end{equation}
Then reference state of the chain can be explicitly written as 
\begin{equation}
\left|\mathcal{G} \right\rangle=\frac{1}{\sqrt{\mathcal{N}}} \prod_{k>0}\left(u_k-v_k \hat{c}_k^{\dagger} \hat{c}_{-k}^{\dagger}\right)|0\rangle,
\end{equation}
with $\mathcal{N}=\prod_{k>0}(\left|u_k\right|^2+\left|v_k\right|^2)$ being the normalization constant.

 \begin{figure*}[t]
\centering
\includegraphics[width=1\textwidth]{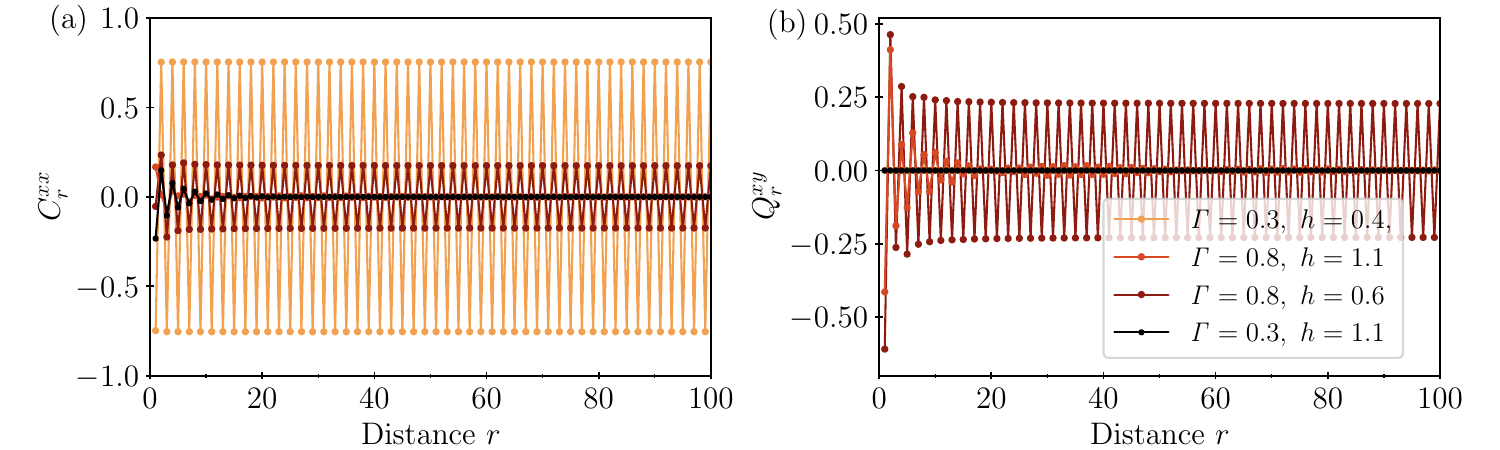}
\caption{(a) The spin-$xx$ correlation function $C_{r}^{xx}$ defined in Eq. (\ref{Cxx}) as a function of distance $r$ for different $\mathit{\Gamma}$ and $h$ [see the legend in (b)]. The results suggest that in both the PT-symmetric and broken phases, the chain exhibits long-range antiferromagnetic order when $h<1$, while the correlation $C_{r}^{xx}$ decays rapidly as $r$ increases when $h>1$. (b) The spin-nematic correlation $Q_{r}^{xy}$ defined in Eq. (\ref{Qxy}) as a function of distance $r$ for different $\mathit{\Gamma}$ and $h$. This shows that the spin-nematic correlation vanishes in the PT-symmetric phase. However, in the PT-broken phase, the chain supports long (short)-range spin-nematic correlations when $h<1$ ($h>1$). Other parameters are taken as $N=2000$ and $J=1$.}
\label{fig_nematic_static}
\end{figure*}

 \begin{figure}[t]
\centering
\includegraphics[width=0.5\textwidth]{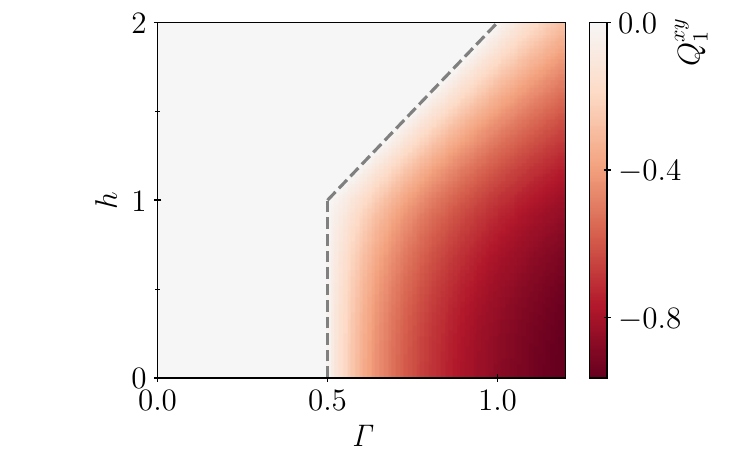}
\caption{The map of local spin-nematic indicator $Q_{1}^{xy}$ in Eq. (\ref{Qxy1}). In comparison with the phase diagram defined as the energy gap in Fig. \ref{fig_gap_phase_diagram}, we see that $Q_{1}^{xy}$ vanishes in the PT-symmetric phase and becomes finite in the PT-broken phase. Other parameters are taken as $N=2000$ and $J=1$.}
\label{fig_nematic_phase_diagram}
\end{figure}

 \begin{figure*}[t]
\centering
\includegraphics[width=1\textwidth]{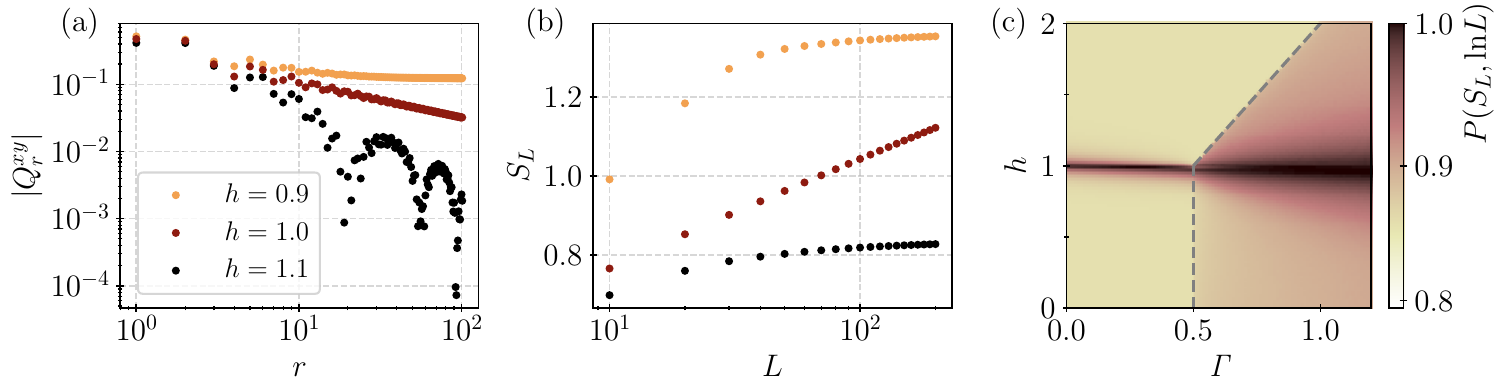}
\caption{(a) Absolute spin-nematic correlation $|Q_{r}^{xy}|$ near $h=1$ in double-logarithmic scale. We observe that $|Q_{r}^{xy}|$ exhibits a power-law decay with distance $r$ at $h=1$. (b) The subsystem entanglement entropy $S_{L}$ as a function of subsystem size $L$ near $h=1$ in semi-logarithmic scale, which demonstrate the logarithmic scaling behavior of the entanglement entropy: $S_{L}\sim \mathrm{ln} L$.  Other parameters are taken as $N=2000$, $J=1$ and $\mathit{\Gamma}=0.8$. (c) The Pearson correlation coefficient $P (S_{L},\mathrm{ln}L)$ between logarithm of the subsystem size $\mathrm{ln} L$ and entanglement entropy $S_{L}$, as a function of $\mathit{\Gamma}$ and $h$. Here $P (S_{L},\mathrm{ln}L)=1$ indicates perfect positive linear correlation $S_{L}\sim \mathrm{ln} L$. We can see that at $h=1$, for all values of $\mathit{\Gamma}$, $S_{L}$ exhibits a linear dependence on $\mathrm{ln} L$. Other parameters are taken as $N=2000$, $J=1$ and the subsystem size  $L$ takes the values $1$, $10$, $20$, ..., $100$.}
\label{fig_EEtropy}
\end{figure*}

In order to clarify the nature of the reference states $\left|\mathcal{G} \right\rangle$ in different phases, we introduce two types of spin-correlation functions. The first one is the spin-$xx$ correlation function
\begin{equation}
	C_r^{xx}=\left< \mathcal{G} \right\vert \hat{\sigma}_{m}^{x} \hat{\sigma}_{n}^{x}\left\vert \mathcal{G} \right>\equiv\left\langle \hat{\sigma}_{m}^{x} \hat{\sigma}_{n}^{x} \right\rangle,
\end{equation}
where $r=n-m$.  Here, the correlation is evaluated using the conventional normalized right-state expectation value, which is associated with the conditional no-jump evolution introduced in \ref{appendix_a}. The resulting correlation functions then directly characterize the spatial structure of the chosen reference state. Moreover, whereas the biorthogonal basis serves mainly as a mathematical framework in non-Hermitian physics, right-state expectation values are more directly connected to experimentally measurable observables. Next, we use the Jordan-Wigner transformation in Eq. (\ref{J-W_transformation}) and write $C_r^{xx}$ as
\begin{equation}
\begin{aligned}
	C_r^{xx}=&\left\langle \left(\hat{c}_m^{\dagger}+\hat{c}_m\right) \prod_{l=m}^{n-1}\left(1-2 \hat{c}_l^{\dagger} \hat{c}_l\right)\left(\hat{c}_n^{\dagger}+\hat{c}_n\right) \right\rangle \\
	=&\left\langle \hat{B}_m \hat{A}_{m+1} \hat{B}_{m+1} \cdots \hat{A}_{n-1} \hat{B}_{n-1} \hat{A}_n\right\rangle.
	\end{aligned}
	\label{Cxx}
\end{equation}
Here we introduce the notation $\hat{A}_{m}=\hat{c}_{m}^{\dagger}+\hat{c}_{m}$ and $\hat{B}_{m}=\hat{c}_{m}^{\dagger}-\hat{c}_{m}$.
With the aid of Wick's theorem \cite{wick1950evaluation}, the above fermionic multi-point correlation function can be reduced to the sums of products of two-point correlation functions for operators $\hat{A}_{m}$ and $\hat{B}_{m}$, which can be recast as the Pfaffian of a $2r \times 2r$ skew-symmetric matrix \cite{Barouch1971}
\begin{widetext}
\begin{equation}
	C_r^{xx}=\mathrm{Pf} \begin{pmatrix}0&\langle \hat{B}_{m}\hat{A}_{m+1} \rangle&\langle \hat{B}_{m}\hat{B}_{m+1} \rangle&\langle \hat{B}_{m}\hat{A}_{m+2} \rangle&\langle \hat{B}_{m}\hat{B}_{m+2} \rangle&\cdots&\langle \hat{B}_{m}\hat{A}_{n} \rangle\\ 
&0&\langle \hat{A}_{m+1}\hat{B}_{m+1} \rangle&\langle\hat{A}_{m+1}\hat{A}_{m+2} \rangle&\langle\hat{A}_{m+1}\hat{B}_{m+2} \rangle&\cdots&\langle\hat{A}_{m+1}\hat{A}_{n} \rangle\\ 
&&0&\langle \hat{B}_{m+1}\hat{A}_{m+2} \rangle&\langle \hat{B}_{m+1}\hat{B}_{m+2} \rangle&\cdots&\langle \hat{B}_{m+1}\hat{A}_{n} \rangle\\ 
&&&0&\langle \hat{A}_{m+2}\hat{B}_{m+2} \rangle&\cdots&\langle \hat{A}_{m+2}\hat{A}_{n} \rangle\\ 
&&&&\ddots&\ddots&\vdots\\ 
&&&&&0&\langle \hat{B}_{n-1}\hat{A}_{n} \rangle\\ 
&&&&&&0\end{pmatrix}.
\label{Cxx_Pf}
\end{equation}
\end{widetext}
The Pfaffian of a skew-symmetric matrix can be evaluated efficiently numerically by the method of tridiagonalization based on Householder transformations \cite{GonzalezBallestero2011, Wimmer2012}. The key is to calculate the correlation functions $\langle \hat{A}_{m} \hat{A}_{n} \rangle$,  $\langle \hat{A}_{m} \hat{B}_{n} \rangle$,  $\langle \hat{B}_{m} \hat{A}_{n} \rangle$ and  $\langle \hat{B}_{m} \hat{B}_{n} \rangle$. This can be done by using the inverse transformation of $\hat{\eta}_k$ in Eq. (\ref{eta_uv})
\begin{equation}
	\hat{c}_k =  \frac{u_k^{\ast} \hat{\eta}_k - v_k \hat{\eta}_{- k}^{\dagger}}{\left| u_{k} \right|^{2} +\left| v_{k} \right|^{2}},
\end{equation}
the anticommutation relation of $\hat{\eta}_k$
\begin{equation}
	\{\hat{\eta}_k, \hat{\eta}_{k^{\prime}}^{\dagger}\}=(\left|u_k\right|^2+\left|v_k\right|^2)\delta_{k, k^{\prime}},
\end{equation}
and the fact of $\hat{\eta}_{k}\left|\mathrm{G} \right\rangle=0$. As a result, we obtain
\begin{equation}
\langle \hat{A}_{m} \hat{A}_{n} \rangle =\delta_{m,n} +\frac{2\mathrm{i}}{\pi} \int_{0}^{\pi} \sin (kr)\frac{\operatorname{Im} \left( u_{k}v_{k}^{\ast} \right)}{\left| u_{k} \right|^{2} +\left| v_{k} \right|^{2}} \mathrm{d} k,
\end{equation}
\begin{equation}
\begin{aligned}
\langle \hat{A}_{m} \hat{B}_{n} \rangle =&\frac{2}{\pi} \int_{0}^{\pi} \sin (kr)\frac{\operatorname{Re} \left( u_{k}v_{k}^{\ast} \right)}{\left| u_{k} \right|^{2} +\left| v_{k} \right|^{2}} \mathrm{d} k \\
&+\frac{1}{\pi} \int_{0}^{\pi} \cos (kr)\frac{\left| u_{k} \right|^{2} -\left| v_{k} \right|^{2}}{\left| u_{k} \right|^{2} +\left| v_{k} \right|^{2}} \mathrm{d} k,
	\end{aligned}
\end{equation}
\begin{equation}
	\langle \hat{B}_{m} \hat{A}_{n}\rangle= -\langle \hat{A}_{n} \hat{B}_{m}\rangle,
\end{equation}
and
\begin{equation}
	\langle \hat{B}_{m} \hat{B}_{n} \rangle =\langle \hat{A}_{m} \hat{A}_{n} \rangle-2\delta_{m,n}.
\end{equation}
Note that for the Hermitian chains, $u_k$ and $v_k$ can in general be chosen to be real, thus $\langle \hat{A}_{m} \hat{A}_{n} \rangle$ and $\langle \hat{B}_{m} \hat{B}_{n} \rangle$ vanish for $m\neq n$, and the Pfaffian in Eq. (\ref{Cxx_Pf}) can be reduced to a determinant \cite{Barouch1971}. However, this is not the case for the non-Hermitian chain considered here since $u_k$ becomes complex in the PT-broken phase. 

The numerical results of the spin-$xx$ correlation function $C_{r}^{xx}$ are presented in Fig. \ref{fig_nematic_static}(a) for typical parameters of $(\mathit{\Gamma},\ h)$. It is shown that in both the PT-symmetric and broken phases, the chain supports long-range antiferromagnetic order when $h<1$, while the correlation $C_{r}^{xx}$ decays rapidly as $r$ increases when $h>1$. We also note that the long-range behaviour of $C_{r}^{xx}$ here is smaller than that of the transverse field Ising chain, which is $\lim_{r\rightarrow \infty} C_{r}^{xx}=(1-h^{2}/J^{2})^{1/4}$ \cite{Pfeuty1970, Sachdev1999}. The transition between ordered and disordered phases in the gapless PT-broken region suggests that there is a region showing signatures consistent with quantum criticality, which cannot be characterized by the energy gap.

The second correlation function considered is the spin-nematic correlation $Q_{r}^{xy}$, which characterizes anisotropic correlations of spin fluctuations between neighboring sites. The spin-nematic correlation is defined as 
\begin{equation}
	Q_{r}^{xy}=C_r^{xy}+C_r^{yx},
		\label{Qxy}
\end{equation}
where
\begin{equation}
\begin{aligned}
	C_r^{xy}=&\left\langle \hat{\sigma}_{m}^{x} \hat{\sigma}_{n}^{y} \right\rangle \\
	=&\mathrm{i}\left\langle \hat{B}_m \hat{A}_{m+1} \hat{B}_{m+1} \cdots \hat{A}_{n-1} \hat{B}_{n-1} \hat{B}_n\right\rangle,
	\end{aligned}
\end{equation}
and
\begin{equation}
\begin{aligned}
	C_r^{yx}=&\left\langle \hat{\sigma}_{m}^{y} \hat{\sigma}_{n}^{x} \right\rangle \\
	=&\mathrm{i}\left\langle \hat{A}_m \hat{A}_{m+1} \hat{B}_{m+1} \cdots \hat{A}_{n-1} \hat{B}_{n-1} \hat{A}_n\right\rangle.
	\end{aligned}
\end{equation}
The fermionic multi-point correlation here can be calculated in the same way as $C_r^{xx}$ in Eqs. (\ref{Cxx}) and (\ref{Cxx_Pf}) . 

In Fig. \ref{fig_nematic_static}(b), we presented the results of spin-nematic correlation $Q_{r}^{xy}$ for typical parameters of $(\mathit{\Gamma},\ h)$. It shows that the spin-nematic correlation vanishes in the PT-symmetric phase, which is different from the spin-$xx$ correlation function. However, in the PT-broken phase, the chain exhibits long-range spin-nematic order when $h<1$ and decays rapidly as $r$ increases when $h>1$. This suggests that the spin-nematic correlation arises from the PT symmetry breaking, and its decay behavior is determined by the spontaneous parity symmetry breaking. It is worth noting that a gapless phase of Hermitian spin chain lacks true long-range order, featuring correlation functions that follow a power-law decay to form a unique quasi-long-range order \cite{haldane1981luttinger, giamarchi2004quantum}. Nevertheless, the non-Hermitian gapless phase here exhibits long-range spin-$xx$ and spin-nematic correlations when $h<1$.

For a more complete picture, we compute the local spin-nematic indicator $Q_{1}^{xy}$ \cite{Yang2021, Soerensen2021, Abbasi2025} in the $\mathit{\Gamma}$-$h$ plane, which is the nearest-neighbor spin-nematic correlation with $r=1$, and provides a simple and sensitive local diagnostic of spin-nematic correlations. The local spin-nematic indicator can be written explicitly as 
\begin{equation}
		Q_{1}^{xy} =   - \frac{4}{\pi} \int_0^{\pi} \sin k 
  \frac{\textrm{\ensuremath{\operatorname{Im}}} (u_k v_k^{\ast})}{| u_k |^2 +
  | v_k |^2}  \mathrm{d} k.
  \label{Qxy1}
\end{equation}
The above expression implies that the spin-nematic indicator vanishes in the PT-symmetric phase since both $u_k$ and $v_k$ are real. Upon increasing the strength non-Hermitian Gamma interaction $\mathit{\Gamma}$ across the critical line, $u_k$ becomes complex due to PT symmetry breaking. As a result, the spin-nematic indicator becomes finite. In the large $\mathit{\Gamma}$ limit, we have $Q_{1}^{xy}= - 4/\pi$. 
Figure \ref{fig_nematic_phase_diagram} presents the results of $Q_{1}^{xy}$ in the $\mathit{\Gamma}$-$h$ plane, which is consistent with our expectation. We would like to point out that the map of $Q_1^{xy}$ should be understood as identifying the region with enhanced local spin-nematic correlations, rather than establishing long-range spin-nematic order throughout the full PT-broken regime. Distinguishing short-range from long-range spin-nematic correlations requires combining the results in Fig. \ref{fig_nematic_static} and the analysis in the following.

\subsection{Subsystem entanglement entropy}

Although in the gapless phase, we cannot identify quantum criticality from the energy $\mathcal{E}$, it can be shown that the state $ |\mathcal{G}\rangle$ displays some signatures consistent with quantum criticality near $h=1$ in terms of  correlation functions and subsystem entanglement entropy. 

First, we compute the absolute spin-nematic correlation $|Q_{r}^{xy}|$ near $h=1$ in double-logarithmic scale at $\mathit{\Gamma}=0.8$, and present the numerical  results in Fig. \ref{fig_EEtropy}(a).
We fit the correlation data for the case of $h=1.0$ in range $r=[10,\ 100]$ to $\left|Q_r^{xy}\right| = A r^{-\eta}$, and find $\eta=0.5164 \pm 0.0052$ and $A=0.3465\pm 0.0071$. This indicate that $|Q_{r}^{xy}|$ exhibits a power-law decay with distance $r$ at $h=1.0$. This scale invariance of the correlation function $Q_{r}^{xy}$ provides a signature consistent with quantum criticality.

 \begin{figure*}[t]
\centering
\includegraphics[width=1\textwidth]{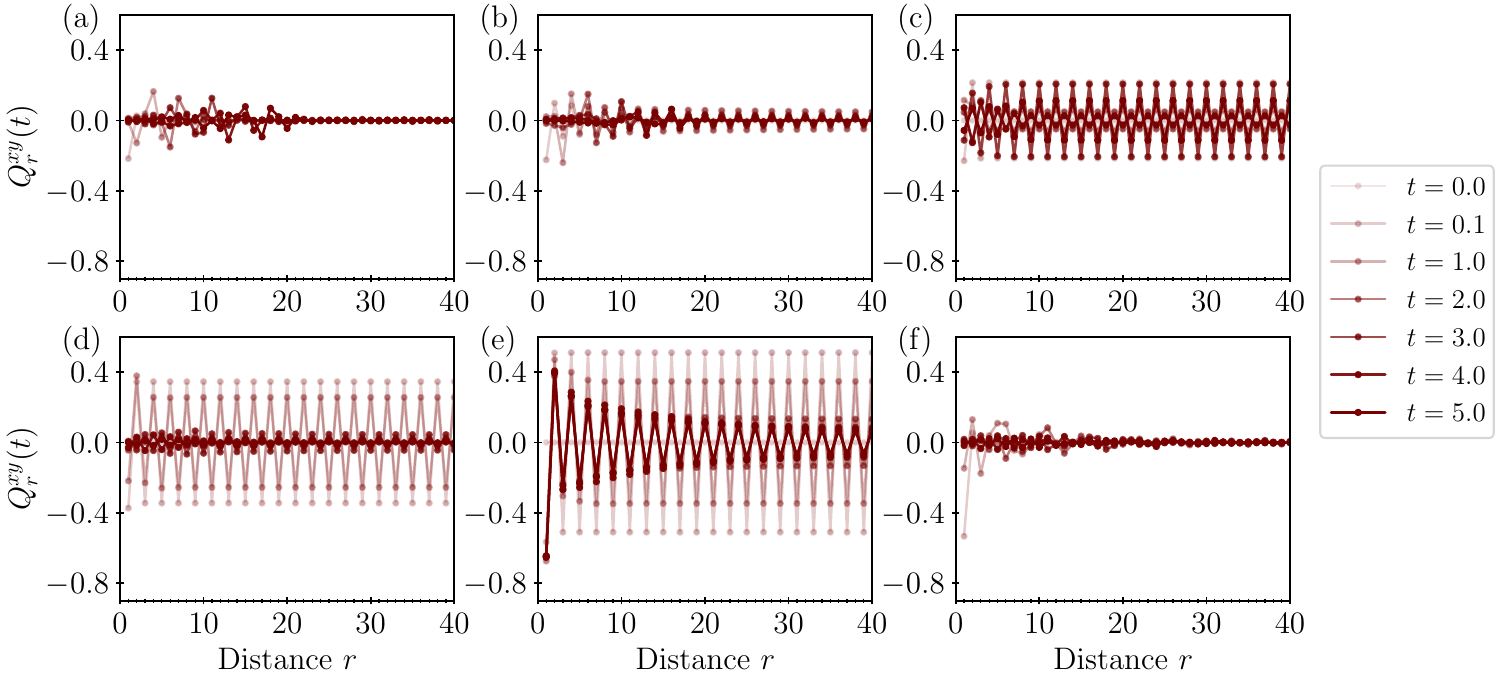}
\caption{(a)-(f) Time evolutions of spin-nematic correlations as a function of distance $r$ at different instants $t$, for the post-quench Hamiltonian  $\mathcal{H}_{k}^{(\mathrm{p})}$ with representative parameters $(\mathit{\Gamma},\ h)$ taken as the same as that in Fig. \ref{fig_gap_phase_diagram} (a)-(f). That is (a) $(\mathit{\Gamma},\ h)=(0.3,\ 1.5)$ in gapped phase; (b) $(0.3,\ 1.0)$ in phase boundary with a degenerate point; (c) $(0.3,\ 0.4)$ in gapped phase; (d) $(0.5,\ 0.4)$ in phase boundary with two exceptional points; (e) $(0.8,\ 0.4)$ in gapless phase with four exceptional points; (f)  $(0.8,\ 1.6)$ in phase boundary with two exceptional points. The pre-quench Hamiltonians have the same transverse fields $h$ as that of the corresponding post-quench Hamiltonians, but $\mathit{\Gamma}=0$. Other parameters are taken as $N=2000$ and $J=1$. In all cases except the PT-broken one in (e), the spin-nematic correlations change signs repeatedly for each $r$ during the time evolutions.}
\label{fig_Energy_Nematic_t}
\end{figure*}

\begin{figure*}[hbt]
\centering
  \includegraphics[width=1\textwidth]{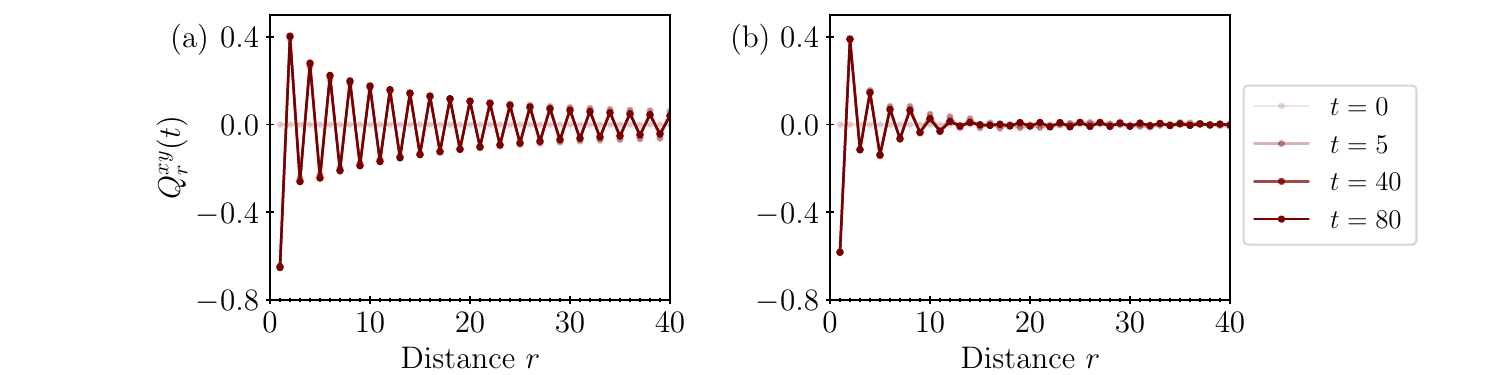}
  \caption{Time evolutions of spin-nematic correlations as a function of distance $r$ at longer instants $t$, for the post-quench Hamiltonian  $\mathcal{H}_{k}^{(\mathrm{p})}$ in the PT-broken region with parameters (a) $\mathit{\Gamma}=0.8,\ h=0.4$ and (b) $\mathit{\Gamma}=1.0,\ h=1.2$. The pre-quench Hamiltonians have the same transverse fields $h$ as that of the corresponding post-quench Hamiltonians, but $\mathit{\Gamma}=0$. Other parameters are taken as $N=2000$ and $J=1$. This indicates that, in the PT-broken region, the spin-nematic correlation $Q_{r}^{xy}(t)$ exhibits robust stability at long times.}
  \label{fig_Nematic_long_t}
\end{figure*}

Second, we consider the subsystem entanglement entropy of state $ |\mathcal{G}\rangle$.  For a subsystem $\mathcal{A}$ of size $L$ and its complement $\mathcal{B}$, the subsystem entanglement entropy is defined as 
\begin{equation}
	S_{L} = -\mathrm{Tr}\left(\hat{\rho}_{L}\ln \hat{\rho}_{L}\right),
\end{equation}
where 
\begin{equation}
	\hat{\rho}_{L}=\mathrm{Tr}_\mathcal{B} \left( |\mathcal{G}\rangle \langle \mathcal{G}| \right)
\end{equation}
is the reduced density matrix of the subsystem $\mathcal{A}$.  In Fig. \ref{fig_EEtropy}(b), we presented the numerical results of $S_{L}$ as a function of subsystem size $L$ near $h=1$ at $\mathit{\Gamma}=0.8$ in semi-logarithmic scale.  We fit the entanglement entropy data for the case of $h=1.0$ in range $L=[10,\ 200]$ to $S_L = A_S \ln L + B_S$ , and find $A_{S}=0.1178\pm 0.0005$ and $B_{S}=0.4996\pm 0.0021$.
This demonstrate the logarithmic scaling behavior of the entanglement entropy, that is, $S_{L}\sim \mathrm{ln} L$, which also provides a signature  consistent with quantum criticality.
The degree of linear correlation between $S_{L}$ and $\mathrm{ln} L$ can be quantified by the Pearson correlation coefficient (PCC). The PCC is defined as 
\begin{equation}
	P (S_{L},\mathrm{ln}L)=\frac{\mathrm{cov}(\mathrm{ln}L, S_{L})}{\sigma_{\mathrm{ln}L} \sigma_{S_{L}}}.
\end{equation}
Here,  $\mathrm{cov}(\mathrm{ln}L, S_{L})$ denotes the covariance between $\mathrm{ln}L$ and $S_{L}$, and $\sigma_{\mathrm{ln}L}$ and $\sigma_{S_{L}}$ denote their standard deviations, respectively. A PCC of 1 indicates perfect positive linear correlation. In Fig. \ref{fig_EEtropy}(c), we present the numerical results of PCC $P (S_{L},\mathrm{ln}L)$ in the  $\mathit{\Gamma}$-$h$ plane. The calculation is carried out based on data for different subsystem sizes with $L=1,10,20, ..., 100$.  We can see that at $h=1$, for all values of $\mathit{\Gamma}$, $S_{L}$ exhibits a linear dependence on $\mathrm{ln} L$. Therefore, the state $ |\mathcal{G}\rangle$ shows signatures consistent with quantum criticality at $h=1$.

\section{Dynamical nematic correlation}
\label{dynamical_nematics_order}

In this section, we investigate the dynamical behavior of the spin-nematic correlation and local spin-nematic indicator under a quench. This may provide a way for generating spin-nematic correlation in the spin chain or characterizing the spin-nematic phase diagram through non-equilibrium dynamics. We consider the post-quench Hamiltonian with nonzero $\mathit{\Gamma}$ and $h$, and the pre-quench Hamiltonian having the same transverse field $h$ as that of the corresponding post-quench Hamiltonian, but $\mathit{\Gamma}=0$. The initial state is the ground state of the pre-quench Hamiltonian, and the evolved state takes the form
\begin{equation}
	\left| \Psi \left( t \right) \right\rangle =\prod_{k>0} \frac{u_{k}(t)-v_{k}(t)\hat{c}_{k}^{\dagger} \hat{c}_{-k}^{\dagger}}{\sqrt{\left| u_{k}(t) \right|^{2} +\left| v_{k}(t) \right|^{2}}} |0\rangle,
	\label{Psi_t_}
\end{equation}
where $u_{k}(t)$ and $v_{k}(t)$ are determined by the time-dependent BdG equation
\begin{equation}
	\mathrm{i} \frac{\mathrm{d}}{\mathrm{d} t} \binom{u_{k}(t)}{v_{k}(t)} =\mathcal{H}_{k}^{(\mathrm{p})} \binom{u_{k}(t)}{v_{k}(t)},
	\label{BdG_t}
\end{equation}
with $\mathcal H_k^{(\mathrm{p})}=-2\mathcal H_k^T$ (see \ref{appendix_c} for the derivation).

In this situation, the evolved state generally oscillates because of the dynamical phase factor. However, non-Hermitian systems may support nonequilibrium steady states, giving rise to dynamical behavior of quantum correlation and entanglement distinct from that of Hermitian systems \cite{Lee2014, Turkeshi2023}. From the previous section, we know that the PT symmetry breaking of the chain leads to non-trivial spin-nematic correlation and local spin-nematic indicator. Actually, in the PT-broken region, the dispersion $\epsilon_{k}$ becomes imaginary for some $k$ and remains real for other $k$ [see Fig. \ref{fig_gap_phase_diagram}(e)], and so does $u_k$ in Eqs. (\ref{ukvk}) and (\ref{Qxy1}). The imaginary components of the dispersion $\epsilon_{k}$ may suppress the dynamical phase factor originating from the real components in the quench, which would result in non-trivial dynamical spin-nematic correlation. To verify this point, we compute the following spin-nematic correlation for the evolved state
\begin{equation}
	Q_{r}^{xy}\left( t \right)=\left\langle \Psi \left( t \right) \right| \left(\hat{\sigma}_{m}^{x} \hat{\sigma}_{n}^{y}+\hat{\sigma}_{m}^{y} \hat{\sigma}_{n}^{x}\right)\left| \Psi \left( t \right) \right\rangle.
\end{equation}
The calculation procedure is the same as that for the static case in the previous section, except that $u_k$ and $v_k$ are replaced by $u_k (t)$ and $v_k (t)$, which are determined by the time-dependent BdG equation in Eq. (\ref{BdG_t}). 

\begin{figure}[t]
\centering
\includegraphics[width=0.5\textwidth]{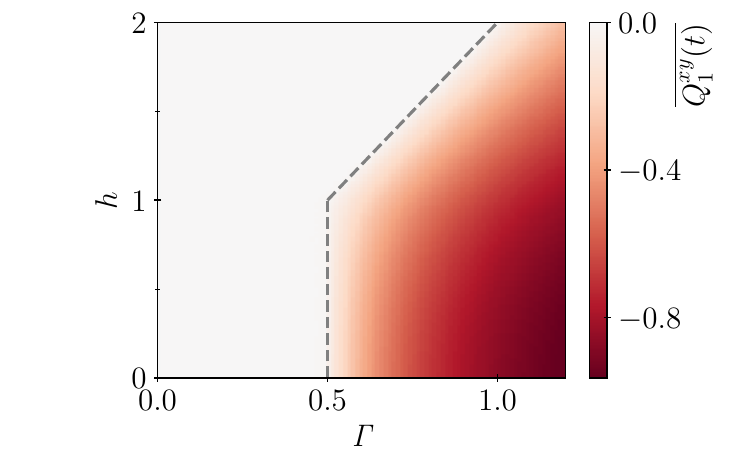}
\caption{Spin-nematic-indicator map obtained from the time average spin-nematic indicator defined in Eq. (\ref{Qxy1t}), for the post-quench Hamiltonian  $\mathcal{H}_{k}^{(\mathrm{p})}$ with different parameters $(\mathit{\Gamma},\ h)$. It is shown that $\overline{Q_1^{x y}(t)}$ captures the non-trivial dynamics of spin-nematic correlation in the PT-broken phase. Here we take $T=50$, $N=2000$ and $J=1$.}
\label{fig_avg_nematic}
\end{figure}

The numerical results of $Q_{r}^{xy}\left( t \right)$ at several different instants $t$ are presented in Fig. \ref{fig_Energy_Nematic_t}. We can see that in all cases except the PT-broken one in Fig. \ref{fig_Energy_Nematic_t}(e), the spin-nematic correlations change signs repeatedly for each $r$ during the time evolutions. The oscillations of $Q_{r}^{xy}\left( t \right)$ due to the dynamical phase factor in the PT-symmetric phase and in the critical lines will lead to its time average vanishing. However, in the PT-broken region, the spin-nematic correlation $Q_{r}^{xy}(t)$ exhibits robust stability at long times, and $Q_{r}^{xy}(t)$ decays more rapidly with distance when $h>1$ (see Fig. \ref{fig_Nematic_long_t}). Nevertheless, it is not a genuine dynamical long-range order, but a robust and persistent dynamical spin-nematic correlations after the quench.
The time average of spin-nematic indicator $Q_{1}^{xy}\left( t \right)$ in the time interval $[0,\ T]$ is 
\begin{equation}
	\overline{Q_1^{x y}(t)}=\frac{1}{T}\int_{0}^{T}Q_{1}^{xy}\left( t \right) \mathrm{d}t.
	\label{Qxy1t}
\end{equation}
This is a dynamical diagnostic that directly measures the stable local spin-nematic component generated during the quench dynamics.  In Fig. \ref{fig_avg_nematic}, we show the numerical results of  the time average spin-nematic indicator $\overline{Q_1^{x y}(t)}$ in the $\mathit{\Gamma}$-$h$ plane, which reproduces the map characterized by the static spin-nematic indicator $Q_{1}^{xy}$ in Fig. \ref{fig_nematic_phase_diagram}.  These results suggest a way for characterizing the spin-nematic map through non-equilibrium dynamics as well as generating spin-nematic correlation in a spin chain.

\section{Conclusion}
\label{conclusion}
In summary, we have demonstrated that in a transverse field Ising chain with non-Hermitian Gamma interaction, the traditional critical point of Ising-type phase transition extend into a critical line, and there is a gapless phase induced by PT symmetry breaking, where the system exhibits long-range and short-range spin-nematic correlations when $h < 1$ and $h > 1$, respectively. This stands in stark contrast to the absence of true long-range order in the gapless phases of Hermitian spin chains. In this gapless phase, the behavior of the spin-$xx$, spin-nematic correlation functions and subsystem entanglement entropy suggest that there are signatures consistent with quantum criticality at $h=1$, which cannot be characterized by the energy gap. Furthermore, we have shown that the PT symmetry breaking leads to the emergence of dynamical spin-nematic correlation,  which provides a way of characterizing the spin-nematic map through non-equilibrium dynamics.  These results show rich quantum phases stem from the competition among the Ising interaction, transverse field and non-Hermitian Gamma interaction, as well as providing a scheme for generating spin-nematic correlation in the spin chain.

\acknowledgments This work was supported by the National Natural Science Foundation of China (Grants No. 12505015, No. 12305026),  the Guangdong Basic and Applied Basic Research Foundation (Grants No. 2024A1515110222), the Research Fund of Guangdong-HongKong-Macao Joint Laboratory for Intelligent Micro-Nano Optoelectronic Technology (No. 2020B1212030010), and the Science \& Technology Development Fund of Tianjin Education Commission for Higher Education(No. 2024KJ060).


\label{A} \setcounter{equation}{0} \renewcommand{\theequation}{A%
\arabic{equation}}
\label{A} \setcounter{figure}{0} \renewcommand{\thefigure}{A%
\arabic{figure}}

\setcounter{section}{0} \renewcommand{\thesection}{APPENDIX A}
\section{Non-Hermitian effective Hamiltonian}
\label{appendix_a}
When the system is coupled to the environment, the dynamics of the state are no longer governed exclusively by the Schrödinger equation, but instead by the Lindblad master equation
\begin{equation}
  \frac{\mathrm{d}}{\mathrm{d} t} \hat{\rho} =  - \mathrm{i} [\hat{H}, \hat{\rho}] + \sum_{j, \mu} \left( \hat{L}_{j, \mu} \hat{\rho}
  \hat{L}^{\dagger}_{j, \mu} - \frac{1}{2} \{  \hat{L}_{j, \mu}^{\dagger}  \hat{L}_{j, \mu},  \hat{\rho}
  \} \right).
\end{equation}
Here the system is the transverse field Ising chain
\begin{equation}
	\hat{H} = J \sum^N_{j = 1} \hat{\sigma}_j^x \hat{\sigma}_{j + 1}^x + h \sum^N_{j = 1}\hat{\sigma}_j^z,
\end{equation}
and the considered jump operators for different dissipative channels are
\begin{equation}
	\begin{aligned}
   \hat{L}_{j, 1} = & \sqrt{\mathit{\Gamma}} ( \hat{\sigma}_j^x +  \hat{\sigma}_{j + 1}^y),\\
   \hat{L}_{j, 2} = & \sqrt{\mathit{\Gamma}} ( \hat{\sigma}_j^y +  \hat{\sigma}_{j + 1}^x),
	\end{aligned}
	\label{jump_op}
\end{equation}
which are correlated dissipations acting on two neighboring lattice sites.
For the quantum trajectory description, we separate the Lindblad master equation into the no-jump evolution and the jump processes
\begin{equation}
  \frac{\mathrm{d}}{\mathrm{d} t}  \hat{\rho} =  - \mathrm{i} ( \hat{\mathcal{H}}  \hat{\rho} -  \hat{\rho}  \hat{\mathcal{H}}^{\dagger}) +
  \sum_{j, \mu}  \hat{L}_{j, \mu}  \hat{\rho}  \hat{L}^{\dagger}_{j, \mu},
\end{equation}
where the non-Hermitian effective Hamiltonian is
\begin{equation}
  \hat{\mathcal{H}} =   \hat{H} - \frac{\mathrm{i}}{2} \sum_{j, \mu}  \hat{L}^{\dagger}_{j, \mu}  \hat{L}_{j,
  \mu},
\end{equation}
Up to a constant term, the non-Hermitian effective Hamiltonian takes the form
\begin{equation}
  \hat{\mathcal{H}} =\hat{H}- \mathrm{i} \mathit{\Gamma} \sum^N_{j = 1} (\hat{\sigma}_j^x \hat{\sigma}_{j + 1}^y + \hat{\sigma}_j^y \hat{\sigma}_{j + 1}^x ),
\end{equation}
which describe the system with no-click measurement outcomes or post-selection. We can see that the imaginary symmetric off-diagonal Gamma interaction occurs in the no-click limit of the stochastic quantum jump trajectories when correlated jump operators in Eq. (\ref{jump_op}) are measured, and $\mathit{\Gamma}\geqslant 0$ is the dissipation rate.

\label{B} \setcounter{equation}{0} \renewcommand{\theequation}{B%
\arabic{equation}}
\label{B} \setcounter{figure}{0} \renewcommand{\thefigure}{B%
\arabic{figure}}

\setcounter{section}{0} \renewcommand{\thesection}{APPENDIX B}
\section{Fermionic Hamiltonian}
\label{appendix_b}

The non-Hermitian Hamiltonian in Eq. (\ref{H_NH}) can be solved by free fermion technique. Through the Jordan-Wigner transformation in Eq. (\ref{J-W_transformation}), the Hamiltonian can be written in quadratic fermionic form 
\begin{equation}
	\begin{aligned}
  \mathcal{\hat{H}} = & J \sum^{N - 1}_{j = 1} (\hat{c}_j^{\dagger} \hat{c}_{j + 1}^{\dagger} -
  \hat{c}_j \hat{c}_{j + 1} + \hat{c}_j^{\dagger} \hat{c}_{j + 1} - \hat{c}_j \hat{c}_{j + 1}^{\dagger}) \\
  & + h\sum^N_{j = 1} (1 - 2 \hat{c}_j^{\dagger} \hat{c}_j)+ 2\mathit{\Gamma} \sum^{N - 1}_{j = 1} ( \hat{c}_j^{\dagger} \hat{c}_{j + 1}^{\dagger} + 
  \hat{c}_j \hat{c}_{j + 1})\\
  & + (- 1)^{\hat{\mathcal{N}} + 1} J (\hat{c}_L^{\dagger} \hat{c}_1^{\dagger} - \hat{c}_L \hat{c}_1 +
  \hat{c}_L^{\dagger} \hat{c}_1 - \hat{c}_L \hat{c}_1^{\dagger})\\
  & + (- 1)^{\hat{\mathcal{N}} + 1} 2\mathit{\Gamma} ( \hat{c}_L^{\dagger} \hat{c}_1^{\dagger} + \hat{c}_L \hat{c}_1),
	\end{aligned}
\end{equation}
where $\hat{\mathcal{N}} = \sum_{j = 1}^N \hat{c}_j^{\dagger} \hat{c}_j$ is the particle-number operator. Clearly, while fermion number is not conserved, its parity $(- 1)^{\hat{\mathcal{N}} + 1}$ is conserved. For an even fermion number, anti-periodic boundary conditions are imposed, whereas for an odd fermion number, periodic boundary conditions are imposed. 
Taking the the Fourier transformation $  \hat{c}_j =  e^{- \mathrm{i} \pi / 4}(N)^{-1/2} \sum_k e^{\mathrm{i} k j} \hat{c}_k$, with the overall factor $e^{- \mathrm{i} \pi / 4}$ being introduced to simplify the notation, the above Hamiltonian can be written as
\begin{equation}
	\begin{aligned}
  \mathcal{\hat{H}} = & J \sum_k (\mathrm{i} e^{\mathrm{i} k} \hat{c}_k^{\dagger} \hat{c}_{- k}^{\dagger} + \mathrm{i} e^{-
  \mathrm{i} k} \hat{c}_k \hat{c}_{- k} + e^{\mathrm{i} k} \hat{c}_k^{\dagger} \hat{c}_k - e^{- \mathrm{i} k} \hat{c}_k \hat{c}_k^{\dagger})\\
  & + h \sum_k (1 - 2 \hat{c}_k^{\dagger} \hat{c}_k)+2 \mathrm{i}\mathit{\Gamma} \sum_k (e^{\mathrm{i} k} \hat{c}_k^{\dagger} \hat{c}_{- k}^{\dagger} + e^{- \mathrm{i} k} \hat{c}_k \hat{c}_{- k}).
	\end{aligned}
	\label{H_k_1}
\end{equation}
The momenta are $k=\pm n\pi/N$ with $n=1,3,5,...,N-1$ for even fermion number sector, and with $n=0,2,4,...,N$ for odd fermion number sector. We are interested in the ground-state energy and correlation functions of the system in the thermodynamic limit, and the effect of the boundary conditions appears only as an $O(1/N)$ correction. Therefore, it suffices to consider the correlation functions within one fermion-parity sector. In this work, we consider the states and dynamics in the even fermion number sector, and take $n=1,3,5,...,N-1$. The Hamiltonian in Eq. (\ref{H_k_1}) can be recast as the BdG form in Eq. (\ref{H_BdG}) by regrouping the terms corresponding to $k$ and $-k$.

\label{C} \setcounter{equation}{0} \renewcommand{\theequation}{C%
\arabic{equation}}
\label{C} \setcounter{figure}{0} \renewcommand{\thefigure}{C%
\arabic{figure}}

\setcounter{section}{0} \renewcommand{\thesection}{APPENDIX C}
\section{Time-dependent B\lowercase{d}G equation}
\label{appendix_c}

Here we present the derivation of the time-dependent BdG equation in Eq. (\ref{BdG_t}). Due to the decoupled nature of the Hamiltonian Eq. (\ref{H_BdG}) and the state Eq. (\ref{Psi_t_}), we can treat each $k$-subspace independently. 
For each $k>0$, the state in Eq. (\ref{Psi_t_}) lies in the even-parity subspace spanned by
\begin{equation}
|0\rangle_k,
\quad
|1\rangle_k=\hat{c}_k^\dagger \hat{c}_{-k}^\dagger |0\rangle_k ,
\end{equation}
where $|0\rangle_k$ is the vacuum state of $(k,-k)$ sector. 
Thus the single-$k$ component of the wave function is
\begin{equation}
|\psi_k(t)\rangle
=
u_k(t)|0\rangle_k
-
v_k(t)|1\rangle_k .
\label{psi_k}
\end{equation}
Writing Eq. (\ref{H_BdG}) as
\begin{equation}
\begin{aligned}
	 \hat{\mathcal{H}} &=  \sum_k \left(\begin{array}{cc}
    \hat{c}_k^{\dagger} & \hat{c}_{- k}
  \end{array}\right) \mathcal{H}_k \left(\begin{array}{c}
    \hat{c}_k\\
    \hat{c}_{- k}^{\dagger}
  \end{array}\right)\\
  &=2\sum_{k>0} \left(\begin{array}{cc}
    \hat{c}_k^{\dagger} & \hat{c}_{- k}
  \end{array}\right) \mathcal{H}_k \left(\begin{array}{c}
    \hat{c}_k\\
    \hat{c}_{- k}^{\dagger}
  \end{array}\right)=\sum_{k>0}\hat{\mathcal{H}}_{k}^{\pm},
  \end{aligned}
  \label{H_BdG_}
\end{equation}
where
\begin{equation}
\mathcal H_k=
\begin{pmatrix}
\varepsilon_k & \Delta_k^+\\
\Delta_k^- & -\varepsilon_k
\end{pmatrix},
\end{equation}
with $\varepsilon_k=J\cos k-h$, $\Delta_k^+=-J\sin k-2\Gamma\sin k$, $\Delta_k^-=-J\sin k+2\Gamma\sin k$, and the summation interval of momenta $k>0$ is $k= n\pi/N$ with $n=1,3,5,...,N-1$. The corresponding second-quantized Hamiltonian in the $(k,-k)$ sector is
\begin{equation}
 \hat{\mathcal{H}}_{k}^{\pm}
=
2\varepsilon_k
\left(
\hat{c}_k^\dagger \hat{c}_k+\hat{c}_{-k}^\dagger \hat{c}_{-k}-1
\right)
+
2\Delta_k^+ \hat{c}_k^\dagger \hat{c}_{-k}^\dagger
+
2\Delta_k^- \hat{c}_{-k}\hat{c}_k .
\end{equation}
Its action on the basis states is
\begin{equation}
 \hat{\mathcal{H}}_{k}^{\pm} |0\rangle_k
=
-2\varepsilon_k|0\rangle_k
+
2\Delta_k^+|1\rangle_k,
\label{action_1}
\end{equation}
and
\begin{equation}
 \hat{\mathcal{H}}_{k}^{\pm} |1\rangle_k
=
2\Delta_k^-|0\rangle_k
+
2\varepsilon_k|1\rangle_k .
\label{action_2}
\end{equation}

Substituting Eqs. (\ref{psi_k}), (\ref{action_1}) and (\ref{action_2}) into the Schr\"odinger equation
\begin{equation}
\mathrm{i} \frac{\partial}{\partial t} |\psi_k(t)\rangle
=
 \hat{\mathcal{H}}_{k}^{\pm} |\psi_k(t)\rangle,
\end{equation}
and comparing the coefficients of $|0\rangle_k$ and $|1\rangle_k$, we obtain
\begin{equation}
\mathrm{i}\frac{\mathrm{d}}{\mathrm{d}t} u_k(t)
=
-2\varepsilon_k u_k(t)-2\Delta_k^- v_k(t),
\end{equation}
and
\begin{equation}
\mathrm{i}\frac{\mathrm{d}}{\mathrm{d}t}v_k(t)
=
-2\Delta_k^+ u_k(t)+2\varepsilon_k v_k(t).
\end{equation}
In matrix form, we have the time-dependent BdG equation
\begin{equation}
\mathrm{i}\frac{\mathrm{d}}{\mathrm{d}t}
\begin{pmatrix}
u_k(t)\\
v_k(t)
\end{pmatrix}
=
\begin{pmatrix}
-2\varepsilon_k & -2\Delta_k^-\\
-2\Delta_k^+ & 2\varepsilon_k
\end{pmatrix}
\begin{pmatrix}
u_k(t)\\
v_k(t)
\end{pmatrix}
\equiv
\mathcal H_k^{(\mathrm{p})}
\begin{pmatrix}
u_k(t)\\
v_k(t)
\end{pmatrix},
\end{equation}
with $\mathcal H_k^{(\mathrm{p})}=-2\mathcal H_k^T$. The normalization factor in Eq. (\ref{Psi_t_}) is imposed after the time evolution of the unnormalized amplitudes and does not modify the linear BdG equation for $u_k(t)$ and $v_k(t)$.

 \begin{figure}[tbh]
\centering
\includegraphics[width=0.5\textwidth]{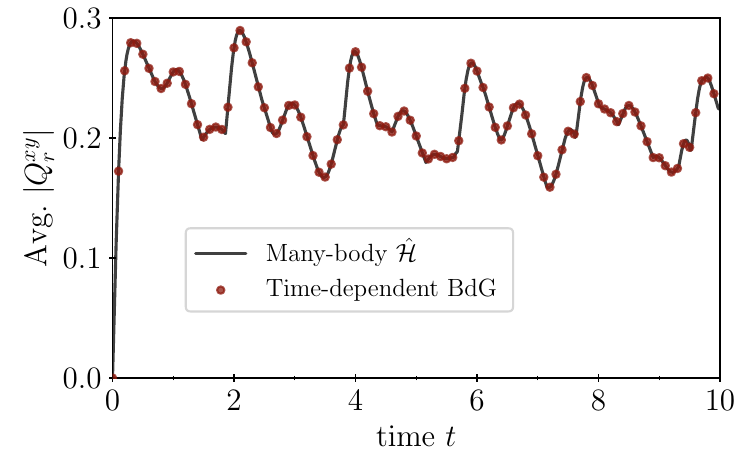}
\caption{Average spin-nematic correlations as a function of time obtain from the time-dependent BdG equation and many-body Hamiltonian. The parameters of the post-quench Hamiltonian are taken as $\mathit{\Gamma}=0.8,\ h=1.2$  and the pre-quench Hamiltonians have the same transverse fields $h$ as that of the post-quench Hamiltonians, but $\mathit{\Gamma}=0$. Other parameters are taken as $N=12$ and $J=1$.}
\label{fig_avg_Q}
\end{figure}

We check the above time-dependent BdG equation and confirm that it is consistent with the many-body Hamiltonian in Eq. (\ref{H_NH}). We evaluate the time evolution of average spin-nematic correlation $\sum_{r=1}^{N-1}|Q_{r}^{xy}|/(N-1)$ for a small system from the two methods. 
The results presented in Fig. \ref{fig_avg_Q} demonstrate that the time-dependent BdG equation in Eq. (\ref{BdG_t}) faithfully reproduces the dynamics generated by the exact diagonalization of the many-body Hamiltonian in Eq. (\ref{H_NH}).

%


\end{document}